\documentclass[twocolumn]{aastex62}
\usepackage[]{amsmath}

\graphicspath{{./}{figs/}}

\received{-}
\revised{-}
\accepted{-}
\submitjournal{ApJ}

\shorttitle{The magnetic field decay of radio pulsars}
\shortauthors{Yi Xie \& Shuang-Nan Zhang}

\begin{document}

\title{The effective magnetic field decay of radio pulsars: insights from the statistical properties of their spin frequency's second derivatives}

\correspondingauthor{Yi Xie, ~Shuang-Nan Zhang}
\email{xieyi@jmu.edu.cn,~zhangsn@ihep.ac.cn}

\author[0000-0003-4179-6394]{Yi Xie}
\affil{School of Science, Jimei University, Xiamen 361021, Fujian Province, China}
\affil{National Astronomical Observatories, Chinese Academy Of Sciences, Beijing 100012, China}

\author[0000-0001-7258-770X]{Shuang-Nan Zhang}
\affil{National Astronomical Observatories, Chinese Academy Of Sciences, Beijing 100012, China}
\affil{Key Laboratory for Particle Astrophysics, Institute of High Energy Physics, Beijing 100049, China}
\affil{Department of Modern Physics, College of Physics, University of Chinese Academy of Sciences, Beijing 100049, China}

\begin{abstract}
We present a new method to investigate the effective magnetic field decay of isolated neutron stars, from the analysis of the long-term timing data of a large sample of radio pulsars \citep{2010MNRAS.402.1027H}. There are some differences between the distributions of frequency's second derivatives of the pulsar spins with different effective field decay timescales. Kolmogorov-Smirnov tests are performed to reexamine the consistency of distributions of the simulated and reported data for a series values of decay timescales. We show that the timescale of the effective field decay exceeds $\sim 5~\rm {Myr}$ for pulsars with spin-down age $\tau_{\rm C}< 10^{7}~{\rm yr}$ or $\sim 100~\rm {Myr}$ for pulsars with $10^{7}<\tau_{\rm C}< 10^{9}~{\rm yr}$ in the sample. The result does not depend on any specific theories of the field evolution, the inclination decay or the variation in the moment of inertia. It is also found that the extent of the closed line region of the magnetic field is close to the light cylinder $r_{\rm lc}$, i.e., the corotating radius $r_{\rm c}\approx r_{\rm lc}$ is a good approximation for the observed pulsar population.

\end{abstract}


\section{Introduction}
\label{sec:intro}

The magnetic field is probably one of the most important physical quantities affecting the evolution and the observational behaviours of radio pulsars. The field strength determines the loss rate of rotational energy, the luminosity of pulses, and thus the spin evolution and observability of a pulsar. The primary method used to determine the magnetic field is by measuring each pulsar's spin parameters, which have actually provided a surprising amount of information on the nature of the pulsed radio sources. As such, the knowledge of the spin parameters is particularly valuable in elucidating whether magnetic field decay occurs in isolated neutron stars. Many impressive studies have been done on this issue during the past few decades \citep[e.g.][]{1969ApJ...157.1395O,1970ApJ...160..979G,1982MNRAS.201..503L,1985MNRAS.213..613L,1987A&A...178..143S,1990ApJ...352..222N,
1992A&A...254..198B,1993MNRAS.261..113H,1997MNRAS.289..592L,1997A&A...318..485H,1998ApJ...505..315C,1998MNRAS.298..625T,2001A&A...376..543T,
2002ApJ...565..482G,2006ApJ...643..332F,2015AN....336..831I,2017MNRAS.467.3493J}. Unfortunately, the conclusions of pulsar population investigations have often been conflicting (See e.g. \cite{2006RPPh...69.2631H,2010MNRAS.404.1081R,2011heep.conf...21L} for reviews). The lack of conclusive evidence on magnetic field decay is mainly due to the fact that the true age of a pulsar is unavailable, and the characteristic (spin-down) age $\tau_{\rm C}$ is normally significantly different from its true age \citep[e.g.][]{2011ASPC..451..231Z}. This makes the evolution of the magnetic field remaining as one of the most important unresolved issues of the physics of neutron stars.

\cite{2010MNRAS.402.1027H} studied the timing noise in the residuals of 366 pulsars that had been regularly observed for 10 to 36 years,  and showed that the magnitude of the frequency's second derivatives of the pulsar spins, i.e. $|\ddot\nu|$, is much larger than that caused by magnetic braking of the neutron star, and the numbers of negative and positive $\ddot\nu$ are almost equal in the sample. It had also been noticed that the distributions between the positive and negative signs in $\ddot\nu-\tau_{\rm C}$ diagram show an approximate symmetry.

In this paper we present a new method to investigate the magnetic field decay of radio pulsars, from the analysis of the long-term timing data. We find that the decay timescale can be constrained by measuring the difference between the distributions of $\ddot\nu$ with different decay timescales in $\ddot\nu-\tau_{\rm C}$ diagram. This method can effectively avoid the ``true age problem". The method and its validity are described in section 2, the revealed restrictions on the timescale of effective magnetic field decay are shown in section 3, and the magnetospheric effects are also tested in the section. The physical implications of the decay timescales are discussed in section 4, and the results are summarized and discussed in section 5.

\section{The method}
\label{sec:methodology}

\subsection{The Spin-down Models}
\label{sec:methodology:models1}

A basic model for a pulsar's spin-down is the magnetic dipole radiation model \citep{1968Natur.219..145P,1969ApJ...157.1395O,1970ApJ...160..979G,1975ApJ...196...51R}. The standard dipole (SD) radiation model assumes that the pure magnetic dipole radiation in vacuum as the braking mechanism, i.e.
\begin{equation}\label{dipole}
\dot\nu =-K \nu^3,
\end{equation}
where $\nu$ and $\dot\nu$ are the spin frequency and its first derivative, respectively. The parameter $K=8\pi^2R^6 B^2\sin^2 \theta/3c^3I$ is a constant, $B=3.2\times 10^{19}\sqrt{-\dot\nu/\nu^3}/\sin\theta$ is the effective dipole magnetic field at equator, $R~(\simeq10^6~{\rm cm})$ and $I~(\simeq10^{45}~{\rm g~cm^2})$ are the radius and moment of inertia, respectively. \cite{2001ApJ...561L..85X} improved the model by incorporating the effect of a longitudinal current outflow (relativistic particle wind) powered by a unipolar generator into the rotation energy-loss rate. This effect was confirmed by a few intermittent pulsars, whose rotation slows down faster when the pulsar is on than when it is off \citep[e.g.][]{2006Sci...312..549K}. Further, \cite{2006ApJ...648L..51S} developed a numerical method for evolving time dependent force-free MHD equations and applied it to solving a dynamic pulsar magnetosphere. Similarly, the dynamic magnetosphere (DM) model also included both the dipole radiation and the unipolar generator mechanisms to contribute the total braking torque of an oblique pulsar, and they found a formula that gives a very good fit to the oblique spin-down for all inclinations. The formula have the same form with Eq.(\ref{dipole}) but with a different parameter $K$,
\begin{equation}\label{K}
K=\frac{4\pi^2 B^2 R^6}{c^3I}(1+\sin^2 \theta).
\end{equation}
The inferred effective magnetic field at the magnetic equator is then $B=2.6\times 10^{19}\sqrt{-\dot\nu/\nu^3}(1+\sin^2 \theta)^{-1/2}$, which can be up to $1.7$ times smaller than the estimate from the SD formula.

\subsection{The Frequency's Second Derivatives and The Revised Spin-down Model}
\label{sec:methodology:models2}

Both the SD and the DM model predict that the frequency second derivative of a pulsar spin $\ddot\nu_{\rm SD}=\ddot\nu_{\rm DM}=3\dot\nu^2/\nu >0$. However, it is widely known that the observed $\ddot\nu$ for the majority of pulsars cannot be explained by these model with a constant field strength $B$. Particularly, the recent large-sample analysis showed \citep{2010MNRAS.402.1027H,2012ApJ...757..153Z,2012ApJ...761..102Z} that $|\ddot\nu|\gg \ddot\nu_{\rm SD}$, $\ddot\nu^{+}(\tau_{\rm C})\approx -\ddot\nu^{-}(\tau_{\rm C})$ and $N(\ddot\nu^{+})\approx N(\ddot\nu^{-})$, where $N$ indicates the total number, the superscripts `+' and `-' indicate positive and negative signs of $\ddot\nu$ and $\tau_{\rm C}\equiv -\nu/2\dot\nu$ is the characteristic age of a pulsar. All the pulsars in the sample are shown in the $|\ddot\nu|-\tau_{\rm c}$ diagram in panel (1) of Fig. \ref{figure1}, in which 193 pulsars have $\ddot\nu>0$ and the remainder 173 pulsars have $\ddot\nu<0$.

Following \cite{1988MNRAS.234P..57B}, we re-formulate the braking law of a pulsar as $\dot\nu=-K(t)\nu^3$, which assumes that the SD or DM model is responsible for the {\it instantaneous} spin-down of a pulsar, but $K(t)$ is {\it time-dependent}. Generically and without depending-upon any specific model for the time-dependence, $K(t)$ can be decomposed into a long-term monotonic term plus a short-term perturbation to the monotonic term. Again assuming $R$ and $I$ are constants, the decomposition is equivalent to $B(t)=B_{\rm M}(t)+B_{\rm O}(t)$, where $B_{\rm M}(t)$ is the long-term monotonic component and $B_{\rm O}(t)$ is the short-term perturbation around $B_{\rm M}$. The quasi-periodic oscillation structures, which is widespread in pulsar timing behaviours \citep{2004MNRAS.353.1311H, 2010MNRAS.402.1027H}, can be phenomenologically described by $B_{\rm O}(t)$ \citep{2013IJMPD..2260012Z,2015RAA....15..963X}. One possible source of the perturbation may be their magnetospheric activities, which are known to influence their timing behaviours significantly \citep{2010Sci...329..408L}. Then after some simple algebra, we get
\begin{eqnarray}\label{ddnu}
\ddot\nu&=&3\dot\nu^2/\nu+2\dot\nu\dot B_{\rm M}/B_{\rm M}+2\dot\nu \dot B_{\rm O}/B_{\rm M}\\  \nonumber
&=&\ddot\nu_{\rm SD}+\ddot\nu_{\rm M}+\ddot\nu_{\rm O}
\end{eqnarray}
where $\ddot\nu_{\rm M}=2\dot\nu\dot B_{\rm M}/B_{\rm M}$ and $\ddot\nu_{\rm O}=2\dot\nu\dot B_{\rm O}/B_{\rm M}$. For relatively old pulsars without significant glitch activities, $\dot\nu<0$ and $\dot B_{\rm M}\leqslant 0$, therefore $\ddot\nu_{\rm M}\geqslant 0$. Given the case of a quasi-periodic oscillation, $\ddot\nu_{\rm O}$ has a positive or a negative value with almost equal chances \citep{2012ApJ...757..153Z,2012ApJ...761..102Z}. Observationally, since generically and statistically $\ddot\nu^{+}(\tau_{\rm C})\approx -\ddot\nu^{-}(\tau_{\rm C})$ and $N(\ddot\nu^{+})\approx N(\ddot\nu^{-})$ for large number of pulsars, clearly $\ddot\nu_{\rm O}$ in Eq. (\ref{ddnu}) dominates the observed statistical properties of $\ddot\nu$. On the other hand, $\ddot\nu_{\rm SD}>0$ and $\ddot\nu_{\rm M}>0$ will cause some differences on the distributions of $\ddot\nu$ with different decay timescales, and some asymmetry between the observed $\ddot\nu^{+}$ and $\ddot\nu^{-}$. These effects might in turn provide clues of long-term magnetic field decay of pulsars, since $\ddot\nu_{\rm SD}$ can be calculated from the observed $\nu$ and $\dot\nu$.

\section{Simulations and Tests}
\label{sec:simu}

\subsection{Monte-Carlo Simulations}
\label{sec:simu:simulations}

We assume that the magnetic fields of pulsars in the sample have a typical decay timescale. Define the timescale as $\tau_{B}\equiv -B_{\rm M}/\dot B_{\rm M}$, we have $\ddot\nu_{\rm M}=-2\dot\nu/\tau_{\rm B}$ and
\begin{equation}\label{ddotnu1}
\ddot\nu=\ddot\nu_{\rm O}+\ddot\nu_{\rm SD}-2\dot\nu/\tau_{\rm B}.
\end{equation}
We can simulate the distributions of $\ddot\nu$ with different $\tau_{B}$, as described below. Our strategy is then to search for the typical $\tau_{B}$ that can maximize the p-value of the Kolmogorov-Smirnov test against the hypothesis that the reported distribution is the same as the simulated distribution in $|\ddot\nu|-\tau_{\rm C}$ diagram for pulsars in the sample.

We construct a phenomenological model for the dipole magnetic field evolution of pulsars with a long-term decay modulated by short-term oscillations,
\begin{equation}\label{B evolution}
B(t)=B_d(t)(1+\sum k\sin(\phi+2\pi\frac{t}{T})),
\end{equation}
where $t$ is the pulsar's age, and $k$, $\phi$, $T$ are the amplitude, phase and period of the oscillation,respectively. $B_d(t)=B_0 \exp(-t/\tau_{\rm B})$, in which $B_0$ is the field strength at the age $t_0$. Substituting Equation (\ref{B evolution}) into Equation (\ref{dipole}), we get the differential equation describing the the spin frequency evolution of a pulsar as follows
\begin{equation}\label{braking law2}
\dot\nu=-A B(t)^2 \nu^{3},
\end{equation}
in which $A=\frac{8\pi^2R^6\sin\theta^2}{3c^3I}$ is a constant, $R~(\simeq10^6~{\rm cm})$, $I~(\simeq10^{45}~{\rm g~cm^2})$, and $\theta~(\simeq\pi/2)$ is the radius, moment of inertia, and angle of magnetic inclination of the neutron star, respectively. The constant $A$ and $B(t)$ are actually inseparable in Equation (\ref{braking law2}), which means that the decay timescales of the effective magnetic fields can be attributed not only to the magnetic field evolution, by also the changes of the inclination angle, or the small changes in the moment of inertia.

In order to model the $\ddot\nu$ distributions in $\tau_{\rm C}- |\ddot\nu|$ diagram, we first obtain $\nu(t)$ by integrating the pulsar spin-down law described as Equation (\ref{braking law2}), and the phase
\begin{equation}\label{phase}
\Phi(t)=\int_{t_0}^{t}\nu(t'){\rm d}t'.
\end{equation}
Then, these observable quantities, $\nu$, $\dot{\nu}$ and $\ddot{\nu}$ can be obtained by fitting the phases to the third order of its Taylor expansion over a time span $T_{\rm s}$,
\begin{equation}\label{phase fit}
\Phi(t_i) =\Phi_0 + \nu (t_i-t_0) + \frac{1}{2}\dot \nu (t_i-t_0)^2
+ \frac{1}{6}\ddot\nu (t_i-t_0)^3.
\end{equation}
We thus get $\nu$, $\dot{\nu}$ and $\ddot{\nu}$ from fitting to Equation~(\ref{phase fit}), with a certain time interval of phases
$\Delta T_{\rm int}=10^6~{\rm s}$.

 We assume that $k=10^{-4.6}\sim 10^{-1.9}$ and $T$ follows a uniform random distribution in the range from $0.1\sim 10 ~{\rm yr}$, and the reasons will be shown in the next subsection. It is also assumed that the sample of the phase $\phi$ of the field oscillations uniformly distributed in the range from $0$ to $2\pi$.  Drawing randomly a data set $\{\nu, \dot\nu, T_{\rm s}\}$ from the reported sample space (i.e. from Table 1 of \cite{2010MNRAS.402.1027H}), and calculating a corresponding start time $t_0$, we can obtain a rotation phase set $\{\Phi(t_i)\}$ using Equation (\ref{phase}). Then the values of $\nu$, $\dot\nu$ and $\ddot\nu$ can be obtained by fitting $\{\Phi(t_i)\}$ to Equation (\ref{phase fit}). Hence one has each $\{\tau_{\rm c},|\ddot\nu| \}$. Repeat this procedure for $N$ times, we will have $N$ ($=283$) data points in the $|\ddot\nu|$-$\tau_{\rm c}$ diagram. We exclude all the pulsars which have glitch records from the sample, since a small variation in the moment of inertial due to glitches may have an impact on the overall spin-down evolution, for instance, a change on the amount of superfluid content of the star may impact the braking index (i.e. \cite{2012NatPh...8..787H}), and actually the timing noise of younger pulsars can be mainly attributed to glitch recovery \citep{2010MNRAS.402.1027H}. Meanwhile, millisecond pulsars ($\tau_{\rm C}\gtrsim 10^{9}~{\rm yr}$ or $\nu>100 ~{\rm s^{-1}}$) are also excluded from the sample, since the characteristic age of millisecond pulsars is highly deceptive, and the millisecond period in these systems reflects the recycling spin-up mechanism rather than secular spin-down evolution.

\begin{figure*}
\centering
\includegraphics[width=20.0cm]{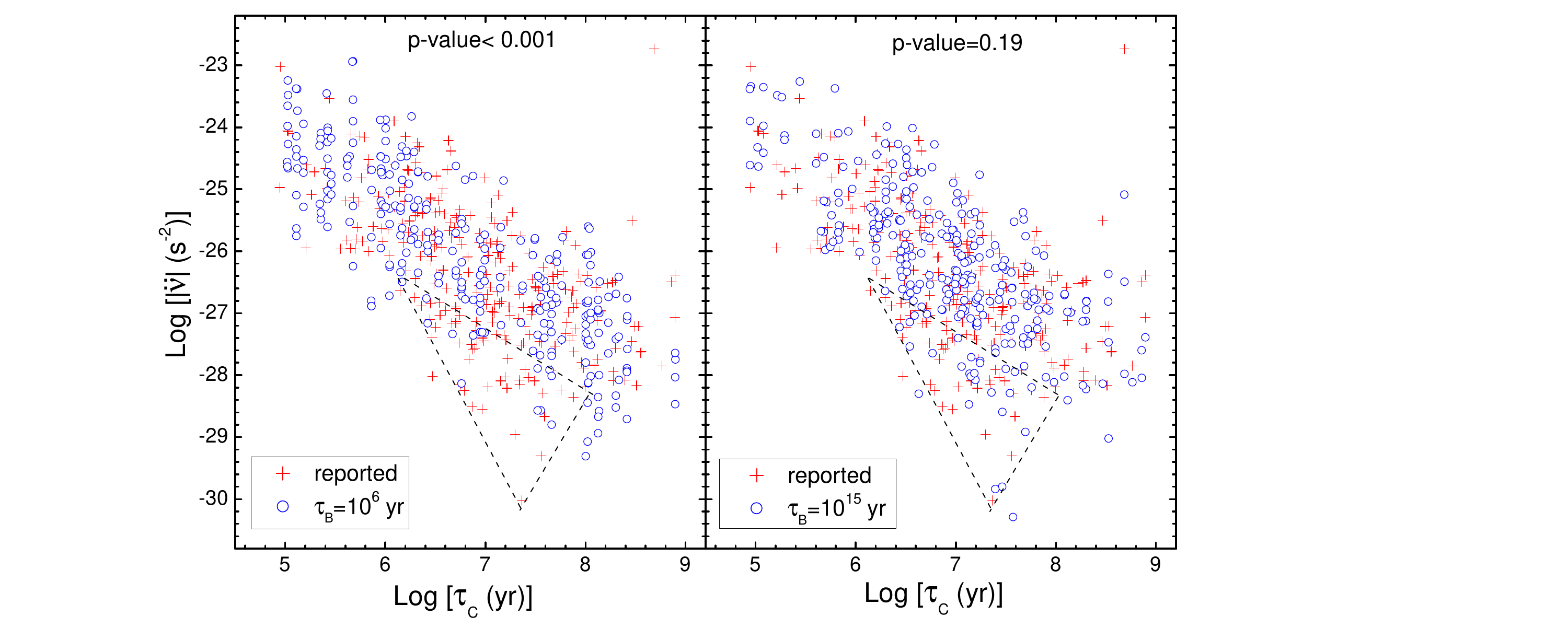}
\caption{$|\ddot\nu|-\tau_{\rm C}$ distributions. The published data from Hobbs et al. (2010) are shown in both panels. The simulated distribution with $\tau_{B}=10^6$ yr is shown in the left panel. The distribution without magnetic decay (with $\tau_{B}=10^{15}$ yr) is shown in the right panel.}
\label{figure1}
\end{figure*}

As examples, we show the simulated $\ddot\nu$ distribution with $\tau_{\rm B}=10^{6}~{\rm yr}$ in left panel of Fig. \ref{figure1}, the simulated distribution without field decay ($\tau_{\rm B}=10^{15}~{\rm yr}$ is taken, which is longer than the age of the universe) in the right panel of Fig. \ref{figure1}, and the reported $\ddot\nu$ distribution in both panels. In the left panel, one can see that the simulated data are much more sparse in the lower part than the reported data, especially inside the triangular area surrounded by dashed lines. In the right panel, the two distributions are completely consistent.

Some of the simulated pulsars in the bottom right part of the sample may have turned off as a radio pulsar and have crossed the death line in the $P-\dot{P}$ diagram. In most models, the period $P$ at turnoff depends upon the structure and the magnitude of the neutron star's surface magnetic field \citep{1993ApJ...402..264C,2000ApJ...531L.135Z}. Assuming a multipole magnetic field configuration, i.e. the polar cap area is similar to that of the pure dipole field, but with very curved field lines at the surface, and the radius of the curvature $r_{\rm c}\sim R=10^6~{\rm cm}$, and this field configuration will be discussed in section \ref{Insight}. The theoretical death line of the pulsar is then \citep{1993ApJ...402..264C},

\begin{equation}\label{deathline}
4 \log B_{\rm p}-6.5 \log P=45.7.
\end{equation}
After taking into account this observational effect, about 30 simulated pulsars have been excluded from the simulated sample in the right panel of Fig. \ref{figure1}.

The analysis on the scatter of $\ddot\nu$ versus $\tau_{\rm C}$ in Fig. \ref{figure1} is potentially misleading, since the two quantities may not be entirely independent. We carry out a similar analysis as Lyne et al. (1975) to assure the reader that inherent correlations could not be found by plotting random pairings of pulsars, i.e. the value of $\nu$, $\dot\nu$, and $\ddot\nu$ are randomly taken from different pulsars in the sample. In this case, there is no clear correlations between $\ddot\nu$ and $\tau_{\rm C}$.

\subsection{Kolmogorov-Smirnov Tests}
\label{sec:simu:test}

\begin{figure*}
\centering
\includegraphics[width=13.0cm]{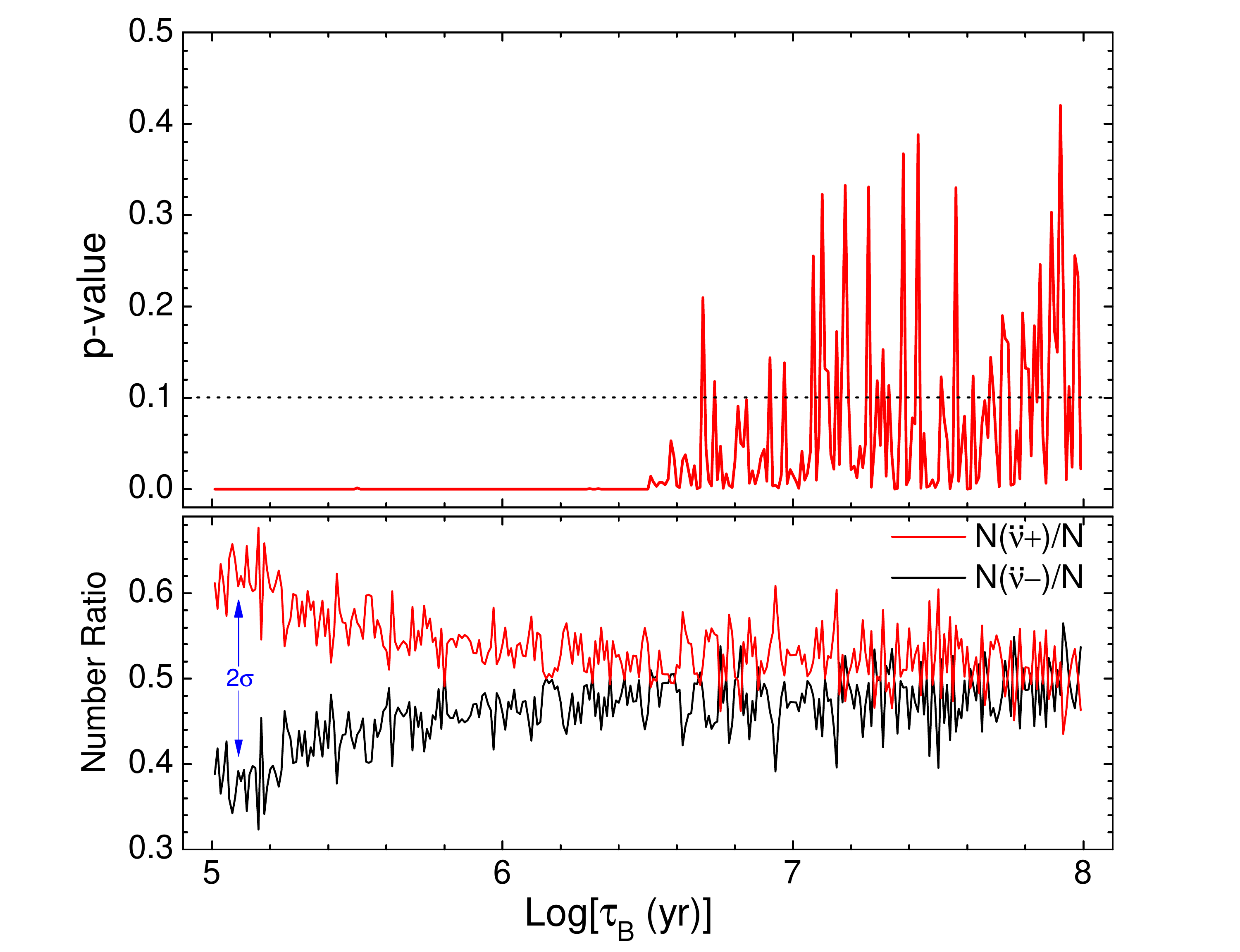}
\caption{Upper panel: the p-values of 2DKS for $|\ddot\nu|-\tau$ distributions from simulated data. The ranges of p-values larger than $0.1$ is identified by the transverse line. The boundary of p-value $\gtrsim 0.1$ indicates $\tau_{B}\gtrsim 5\times 10^{6}~\rm {yr}$. Bottom panel: the pulsar number ratios of the positive and negative $\ddot\nu$ against $\tau_{\rm B}$ are represented by red and black lines, respectively. $\sigma$ is the standard deviation of Poisson distribution.}
\label{figure2}
\end{figure*}

We perform two-dimensional Kolmogorov-Smirnov (2DKS) test to reexamine the consistency of distributions of the simulated and reported $\ddot\nu$ for a series values of $\tau_{B}$. The 2DKS package\footnote{{http://www.downloadplex.com/Scripts/Matlab/Development-Tools/two-sample-two-diensional-kolmogorov-smirnov-test\b{~}432625.html}} \citep{1983MNRAS.202..615P} is adopted for the test. Our strategy is then to search for a typical $\tau_{B}$ that can maximize the p-value of the 2DKS test against the hypothesis that the two distributions are consistent for the pulsars in the sample. We let $\tau_{\rm B}$ vary from $10^5$ to $10^{8}$ yr. The returned p-values are shown with solid lines in the upper panel of Fig. \ref{figure2}. The p-value $0.1$ is considered as the threshold level with probability $95\%$. From the panel, one can see that the decay timescale can be well constrained with p-values larger than $\sim 0.1$. It is shown apparently that the decay timescale $\tau_{B}\gtrsim 5\times 10^{6}~\rm {yr}$. In the bottom panel, we show the $N(\ddot\nu^{+})$ and $N(\ddot\nu^{-})$ as functions of $\tau_{\rm B}$, giving a constraint $\tau_{\rm B}\gtrsim 10^{5}$ yr for $95\%$ probability, which is much loose than the constraint from 2DKS tests. It should be noticed that the 2DKS test has only an approximate and stochastic p-value in each simulation, thus very intensive tests (with $\log\Delta\tau_{B}=0.01$) were performed, as shown in the upper panel. We also checked the validity of 2DKS tests with one-dimensional Kolmogorov-Smirnov (1DKS) test. A very similar result is obtained with 1DKS but $\tau_{B}\gtrsim 10^{6} \rm {yr}$.

\begin{figure*}
\centering
\includegraphics[width=15.0cm]{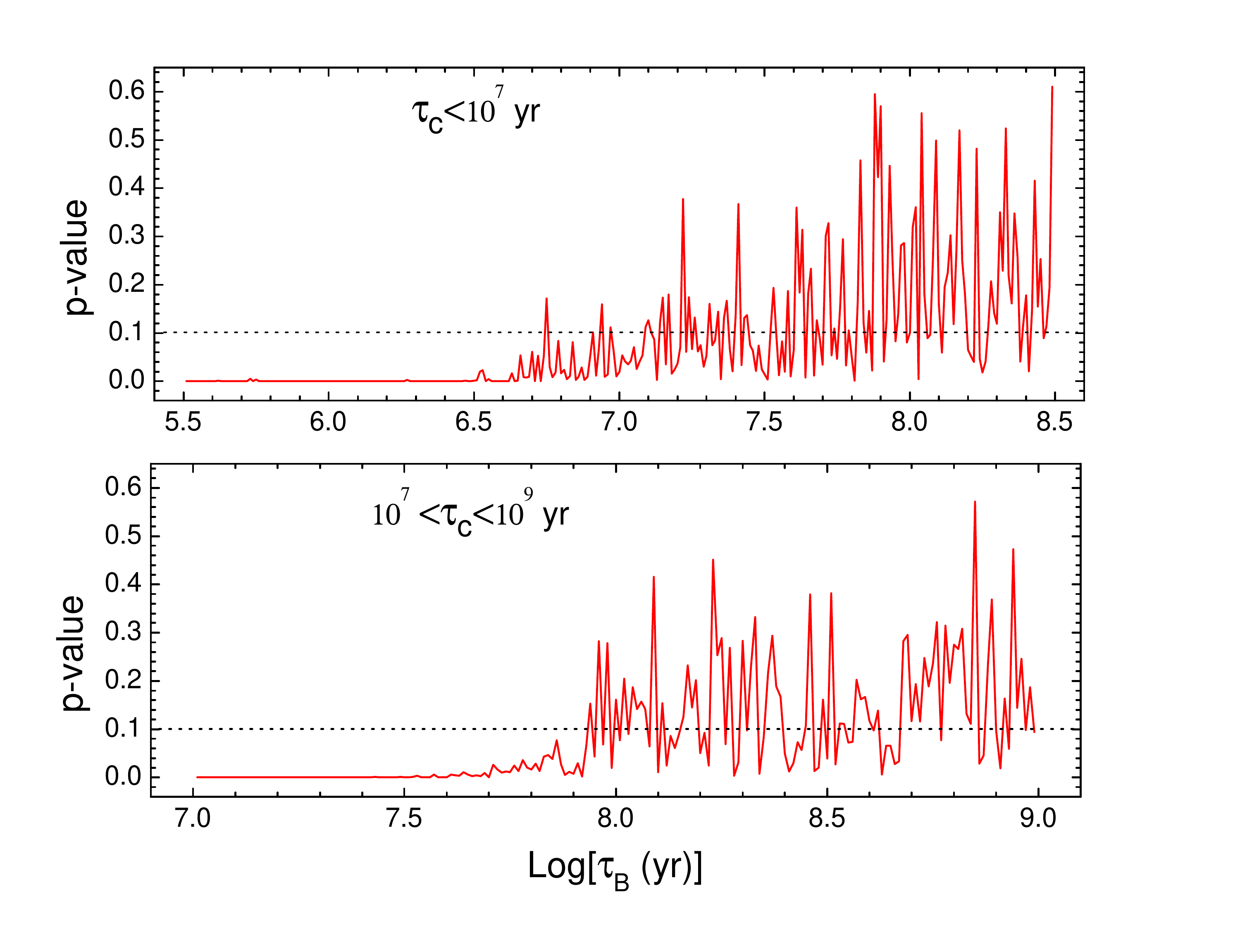}
\caption{Upper panel: the p-values of 2DKS for $|\ddot\nu|-\tau$ distributions from young pulsars with $\tau_{\rm C}< 10^{7}~{\rm yr}$. Bottom panel: the p-values of 2DKS for $|\ddot\nu|-\tau$ distributions from young pulsars with $10^{7}<\tau_{\rm C}< 10^{9}~{\rm yr}$.}
\label{figure3}
\end{figure*}

We also performed the 2DKS test only for young pulsars with $\tau_{\rm C}< 10^{7}~{\rm yr}$, the returned values also give $\tau_{B}\gtrsim 5\times 10^{6}~\rm {yr}$, as shown in the upper panel of Fig. \ref{figure3}. However, for the sample of middle-age pulsars with $10^{7}<\tau_{\rm C}< 10^{9}~{\rm yr}$, the returned values give $\tau_{B}\gtrsim 10^{8}~\rm {yr}$, as shown in the bottom panel of Fig. \ref{figure3}. In addition, 23 millisecond recycled pulsars are excluded from our samples, and the number is too small to be tested independently with 2DKS.

It is very important to explore the parameter space of the simulations, especially regarding the dependence on the oscillation period $T$ and magnitude $k$. However, it is found that the method cannot place effective restrictions on $T$. For instance, there is no significant difference between the returned p-values for $T$ distributing uniformly from $5$ to $25~{\rm yr}$ or from $0.1$ to $100~{\rm yr}$. There are extreme examples of magnetars with torque variations which could be interpreted as a change in the spin-down magnetic field by a generous fraction within months (i.e. \cite{2015ApJ...800...33A} in 1E 1048.1-5937). Thus, for a wider coverage the latter ($0.1\sim 100~{\rm yr}$) is chosen for all the simulations in this paper. For the magnitude $k$, the returned p-values for the lower limit and the upper limit ($k=10^{-k_{2}}\sim 10^{- k_{1}}$) of the power index are shown in panel (a) of Fig. \ref{figure4}. It can easily be seen that $3.9 \lesssim k_{1} \lesssim 4.6$ and $1.9\lesssim k_{2} \lesssim 2.5$. Therefore, $k=10^{-4.6}\sim 10^{-1.9}$ is taken for the wider coverage. As an example, the young pulsar PSR ${\rm B1828 - 11}$ shows correlated shape and spin-down changes  \citep{2019MNRAS.485.3230S}, and the observed $0.7\%$ variation in $\dot P$ implies a fractional change of similar magnitude in the oscillation magnitude $k\simeq 3.5 \times 10^{-3}$, which falls well within the limits.

\subsection{The magnetospheric effects}

Aside from the two prevailing models (SD and DM model), \cite{2006ApJ...643.1139C} proposed a spin-down formula that takes into account the magnetospheric particle acceleration gaps and the misalignment of magnetic and rotation axes, as well as the mechanism of the magnetic field reconnection around the equatorial extent $r_{\rm c}$ of the closed-line region. This formula can be simply expressed as
\begin{equation}\label{dipole3}
\dot\Omega =\frac{B^2 R^6 \Omega}{4cIr_{\rm c}^2}[\sin^2\theta+(1-\frac{\Omega_{\rm death}}{\Omega})\cos^2\theta],
\end{equation}
where the corotating region follows the light cylinder as $r_{\rm c}=r_{\rm lc}(\Omega/\Omega_0)^\alpha$, $\Omega_0$ is the value of the angular velocity $\Omega$ at pulsar birth, and the parameter $\alpha$ ($0<\alpha<1$) depends on the efficiency of the reconnection around $r_{\rm c}$. If the reconnection is very efficient, $r_{\rm c}\approx r_{\rm lc}$, i.e. $\alpha=0$.  However, if the reconnection is very inefficient, then the closed-line region cannot grow, thus $r_{\rm c}\approx {\rm const}.$ \citep{2006ApJ...643.1139C}. $\Omega_{\rm death}$($=2\pi/P_{\rm death}$) describes a pulsar is ``death", i.e., the cessation of pulsar emission. $P_{\rm death}$ can be written as
\begin{equation}\label{deathline}
\begin{split}
P_{\rm death} =& 8.1^{1/(2-\alpha)}~{\rm s}~(\frac{B}{10^{12}~{\rm G}})^{1/(2-\alpha)}\\
& \times (\frac{V_{\rm gap}}{10^{12}~{\rm V}})^{-1/(2-\alpha)} (\frac{P_0}{1~\rm s})^{-1/(2-\alpha)},
\end{split}
\end{equation}
in which $V_{\rm gap}$ is the gap potential.

\begin{figure*}
\centering
\includegraphics[width=18.0cm]{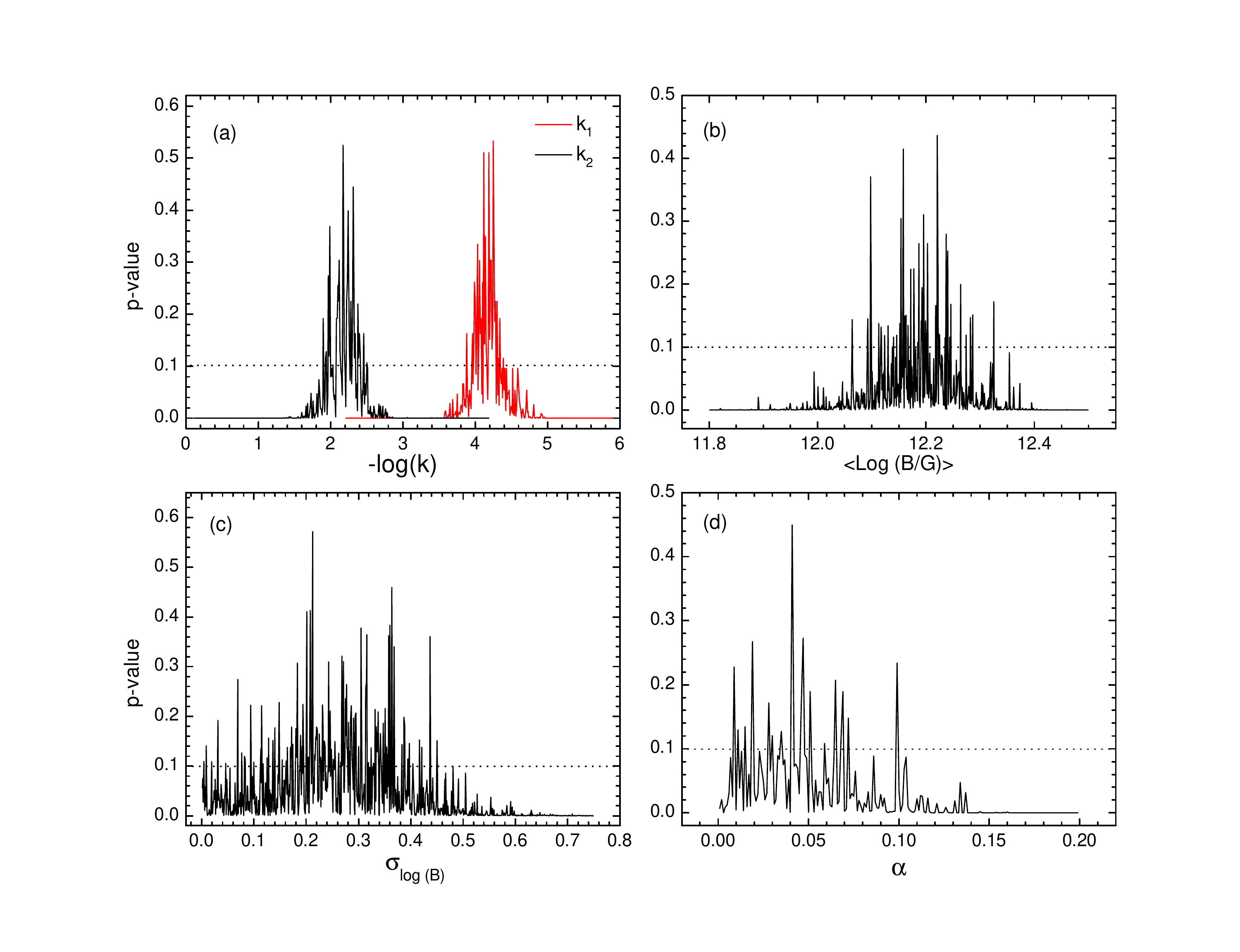}
\caption{The p-values of 2DKS for $|\ddot\nu|-\tau$ distributions from simulated data. The ranges of p-values larger than $0.1$ is identified by the transverse line. Panel (a): the p-values for various values of the upper limit $ k_{1} $ and the lower limit $ k_{2} $ of $-\log k$; Panel (b): the p-values for various values of $\langle\log(B/{\rm G})\rangle$; Panel (c): the p-values for various values of $\sigma_{\log B}$; Panel (d): the p-values for various values of $\alpha$.}

\label{figure4}
\end{figure*}

We perform Monte Carlo simulations to confront the model with observations. The main procedures of the simulations are the same as in the previous subsections. We assume a lognormal distribution of polar magnetic fields with mean value $\langle\log(B/{\rm G})\rangle$ and standard derivation $\sigma_{\log B}$. Following \cite{2006ApJ...643.1139C}, $V_{\rm gap}=10^{13}~{\rm V}$ is taken, and the initial period $P_0$ is uniformly distributed between $10~{\rm ms}$ and $0.2~{\rm s}$. We assume the distribution of inclination angle $\theta$ is also uniform from $0$ to $\pi/2 $. The returned p-values for $\langle\log(B/{\rm G})\rangle$, $\sigma_{\log B}$ and $\alpha$ are shown in the panels (b), (c) and (d) of Fig. \ref{figure4}, respectively. The results show that $12.05\lesssim\langle\log(B/{\rm G})\rangle \lesssim 12.35$, $\sigma_{\log B}\lesssim 0.50 $, and  $\alpha\lesssim 0.11$.

The result of no floor for $\sigma_{\log B}$ implies that the distribution width of $\ddot\nu$ is determined by the oscillation magnitude $k$, rather than by the distribution width of the magnetic field $B$. The parameter $\alpha\lesssim 0.11$ means that our method cannot prove or rule out the spin-down law, but suggests that the reconnection of the north-south poloidal magnetic field around $r_{\rm lc}$ is very efficient, and the extent of the closed line region is close to the light cylinder. We thus propose that $\alpha\sim 0$ or $r_{\rm c}\approx r_{\rm lc}$, is a good approximation for the observed pulsar population.


\section{Physical Implications}
\label{Insight}

The time-dependent behaviors of $B(t)$, and thus the decay timescales of the effective magnetic fields, can be attributed not only to the magnetic field evolution, by also the changes of the inclination angle, or the small changes in the moment of inertia. However, since these tests only prescribe lower limits on the evolution timescales, the results are valid for all the three mechanisms.

Magnetic field are crucial for neutron stars's activities. Understanding the long-term evolution of neutron stars' magnetic fields might be key to unifying the observational diversity of isolated neutron stars \citep{2013MNRAS.434..123V}. The magnetic field in slow-rotating ultra-magnetized neutron stars, so-called magnetars (AXPs and SGRs), is believed to be decay on timescales of $10^3-10^5$ years \citep{1996ApJ...473..322T}, since their rotational energy is not sufficient to power the observed emission. The isolated X-ray pulsars with spin periods longer than $12~\rm s$ are still rarely observed. However, they are not subject to physical limits to the emission mechanism nor observational biases against longer periods. This puzzle could be well understood if their magnetic field is dissipated by one or even two orders of magnitude for $1~{\rm Myr}$, which is probably due to a highly resistive layer in the innermost part of the crust of neutron stars \citep{2013NatPh...9..431P}. For normal radio pulsars, some population synthesis studies suggest that $\tau_{\rm B}$ must be longer than $10~{\rm Myr}$ \citep{1997A&A...322..477H,2001A&A...374..182R}. However, there are also some other studies claimed short decay timescales, i.e. $0.1\lesssim \tau_{\rm B}\lesssim 10~{\rm Myr}$ \citep{1985MNRAS.213..613L,1990ApJ...352..222N,2004ApJ...604..775G,2010MNRAS.401.2675P,2014MNRAS.443.1891G,2014MNRAS.444.1066I}. The present method imposes a piecewise restriction on the decay timescale, i.e. $\tau_{\rm B}\gtrsim 5~{\rm Myr}$ for young pulsars ($\tau_{\rm C}< 10^{7}~{\rm yr}$), and $\tau_{\rm B}\gtrsim 100~{\rm Myr}$ for middle-age pulsars ($10^{7}<\tau_{\rm C}< 10^{9}~{\rm yr}$), and may contribute to our understanding of actual mechanisms of the field decay and magnetic configurations in neutron stars.

Three avenues for the magnetic field decay in isolated neutron stars have been intensively studied, i.e. Ohmic decay, ambipolar diffusion, and Hall drift \citep[e.g.][]{1992ApJ...395..250G,1999MNRAS.304..451U,2002A&A...392.1015G,2002PhRvL..88j1103R,2002MNRAS.337..216H,2004ApJ...609..999C,2007A&A...470..303P,2010A&A...513L..12P,2012A&A...547A...9P,2012MNRAS.421.2722K,2013MNRAS.435.3262G,2014PhRvL.112q1101G}. Depending on the strength of the magnetic fields, each of these processes may dominate the evolution. Ohmic decay occurs in both the fluid core and solid crust. It is inversely proportional to the electric conductivity and independent of the field strength. The Hall drift is non-dissipative and thus cannot be a direct cause of magnetic field decay. However, it can enhance the rate of ohmic dissipation, since only electrons are mobile in the solid crust, and their Hall angle is large. This causes that the evolution of magnetic fields resembles that of vorticity, and then the fields undergo a turbulent cascade terminated by ohmic dissipation at small scales \citep{1992ApJ...395..250G,2004ApJ...609..999C}. Compared with the Hall drift, the timescale of the ambipolar diffusion is much longer for normal pulsars, however, it may be very important for magnetars \citep{1996ApJ...473..322T}. For
a typical density and conductivity profile in the crustal region \citep[e.g.][]{2007A&A...470..303P,2014MNRAS.438.1618G}, the Ohmic timescale is
\begin{equation}\label{ohmic time}
\tau_{\rm Ohm}\sim \frac{4\pi\sigma L^2}{c^2}=13.5(\frac{\sigma}{3\times 10^{24}~{s^{-1}}})(\frac{L}{\rm km})^2~{\rm Myr},
\end{equation}
where $\sigma$ is the electric conductivity, and $L$ is the characteristic length scale of magnetic field in the crust.
For the Hall timescale, one reads,
\begin{equation}\label{hall time}
\tau_{\rm Hall}\sim \frac{4\pi eL^2 n_{\rm e}}{cB}=\frac{16.8}{B_{13}}(\frac{n_{\rm e}}{{2.5\times  10^{36}~{\rm cm}^{-3}}})(\frac{L}{\rm km})^2~{\rm Myr},
\end{equation}
in which $B_{13}\equiv B/(10^{13}~\rm G)$. The combined effect, i.e. Hall cascade, could cause a fast field evolution on a timescale of the order of $10~{\rm Myr}$ \citep{2015MNRAS.453..671G}.

All these contradictory facts can be well understood by the natural assumption that the magnetic field is maintained by two current systems. The large scale dipolar field which is responsible for the pulsar spin down are supported by long living currents in the superconducting core. Currents in the crust support the small scale multipolar fields which decay on timescale that are comparable to the pulsar spin-down ages \citep{2007A&A...470..303P}. The two current systems and the corresponding field configurations are particularly demonstrated in the burst activities of a low dipole magnetic field magnetar, SGR 0418+5729, which is expected to harbor a sufficiently intense internal toroidal component \citep{2010Sci...330..944R}. The present result for middle-age pulsars, i.e. $\tau_{B}\gtrsim 100~\rm {Myr}$, suggests that the dipole component that anchored in the crust are relatively low, and thus its decay has no observable influence on the spin frequency's second derivatives of pulsars in the sample.
In addition, the core-anchored field could be expelled and subsequently dissipated in the crust, and our result also implys that the timescale exceeds $5~\rm {Myr}$. This may be helpful to understand the poorly known physics at the crust-core boundary.

Our results are also suitable for changes of the inclination angle, which could be either due to rotation-magnetic axis alignment or three-dimensional magnetic field evolution \citep{2014MNRAS.441.1879P,2018ApJ...852...21G}. Using polarization data for a large number of isolated pulsars, \cite{1998MNRAS.298..625T} found that the magnetic beam axis align with the spin axis on a timescale of $\sim 10~{\rm Myr}$. With new data, \cite{2010MNRAS.402.1317Y} found a shorter alignment timescale of $\sim 1~{\rm Myr}$. Theoretically, the electromagnetic torque which brakes the rotation of a pulsar also tends to align the magnetic axis with the rotation axis \citep{1970ApJ...159L..81D,1970ApJ...160L..11G}. The electromagnetic alignment timescale is related to the spin-down age as \citep{2018MNRAS.481.4169L}, 
\begin{equation}\label{alignment time}
\tau_{\rm A}\equiv \frac{\sin{\theta}}{\frac{d}{dt}\sin{\theta}} =2\frac{\sin^2{\theta}}{\cos^2{\theta}}\tau_{\rm c}.
\end{equation}
For young or middle-age pulsars, our results imply the alignment timescale is most likely longer than $\sim 5~{\rm Myr}$ or $\sim 100~{\rm Myr}$, which is roughly consistent with the relation.

A small variation in the moment of inertia may have an impact on the overall spin-down evolution, for instance, a decrease in the effective moment of inertia due to an increase on the amount of superfluid content as the star cools through neutrino emission may impact the braking index \citep{2012NatPh...8..787H}. However, most of the stars are typically young and glitching pulsars, which have been excluded from our sample. Our results imply that the populations without glitch record shows no long-term variation in the moment of inertia with timescale short than $5~{\rm Myr}$.



\section{Summary and Discussion}

The perturbation from the long-term dipole magnetic field decay will produce some differences on the distributions for the second derivatives of pulsars' spin frequency with different decay timescales. This in turn provides a new method to investigate the magnetic field decay of radio pulsars, which does not depend on any specific theories of field evolution or inclination decay. We made use of the published large-sample timing data of radio pulsars to find evidence of their magnetic field decay with 2DKS tests. The method impose a piecewise restriction on the decay timescale, i.e. $\tau_{\rm B}\gtrsim 5~{\rm Myr}$ for young pulsars with $\tau_{\rm C}< 10^{7}~{\rm yr}$, and $\tau_{\rm B}\gtrsim 100~{\rm Myr}$ for middle-age pulsars with $10^{7}<\tau_{\rm C}< 10^{9}~{\rm yr}$. It is also proposed that the corotating radius $r_{\rm c}\approx r_{\rm lc}$ is a good approximation for the observed pulsar population. Though pulsars with major glitches have been excluded from the data, tiny glitch activities and other types of timing irregularities may still have some influences on the observed $\ddot\nu$, which may cause a small deviation.

We expect to gain much deeper understanding of pulsars from future larger sample of radio pulsars with higher precision data on $\ddot\nu$, to be brought by China's soon-to-be operating Five-hundred-meter Aperture Spherical radio-Telescope (FAST) and the future Square Kilometer Array (SKA). \\

\section*{Acknowledgments}

We thank J. Y. Liao for discussions. We thank the anonymous referee for comments and suggestions that led to a significant improvement in this manuscript. This work is supported by National Natural Science Foundation of China under grant Nos. 11603009, 11803009, 11373036,and 11133002, by the National Program on Key Research and Development Project under grant Nos. 2016YFA0400802, by the Key Research Program of Frontier Sciences, CAS, Grant No. QYZDY-SSW-SLH008, and by the Natural Science Foundation of Fujian Province under grant Nos. 2016J05013 and 2018J05006.

\end{document}